\documentclass{phb-proc4-auth}

\usepackage{graphicx}
\usepackage{amssymb}
\usepackage[numbers,sort&compress]{natbib}

\begin{document}
\begin{frontmatter}

\journal{SCES '04}

\title{Third-order magnetic susceptibility of the frustrated
square-lattice antiferromagnet}

\author{Burkhard Schmidt\corauthref{1}}
\author{Peter Thalmeier}
 
\address{Max-Planck-Institut f\"ur Chemische Physik fester Stoffe,
Dresden, Germany}

\corauth[1]{Corresponding author: Max-Planck-Institut f\"ur Chemische
Physik fester Stoffe, N\"{o}thnitzer Str.  40, 01187 Dresden, Germany.
Phone: +49 351 4646-2235, fax: +49 351 4646-3232, e-mail:
bs@cpfs.mpg.de}

\begin{abstract}
    We present results from our analysis of the finite-temperature
    properties of the spin 1/2 $J_{1}$-$J_{2}$ Heisenberg model on a
    square lattice.  The analysis is based on the exact
    diagonalization of small clusters with 16 and 20 sites utilizing
    the finite-temperature Lanczos method.  In particular, we focus on
    the temperature dependence of the third-order magnetic
    susceptibility as a method to resolve the ambiguity of exchange
    constants.  We discuss the entire range of the frustration angle
    $\phi=\tan^{-1}(J_{2}/J_{1})$ parameterizing the different
    possible phases of the model, including the large region in the
    phase diagram with at least one ferromagnetic exchange constant.
\end{abstract}


\end{frontmatter}

\begin{figure}[t]
    \centering
    \includegraphics[width=.496\columnwidth]{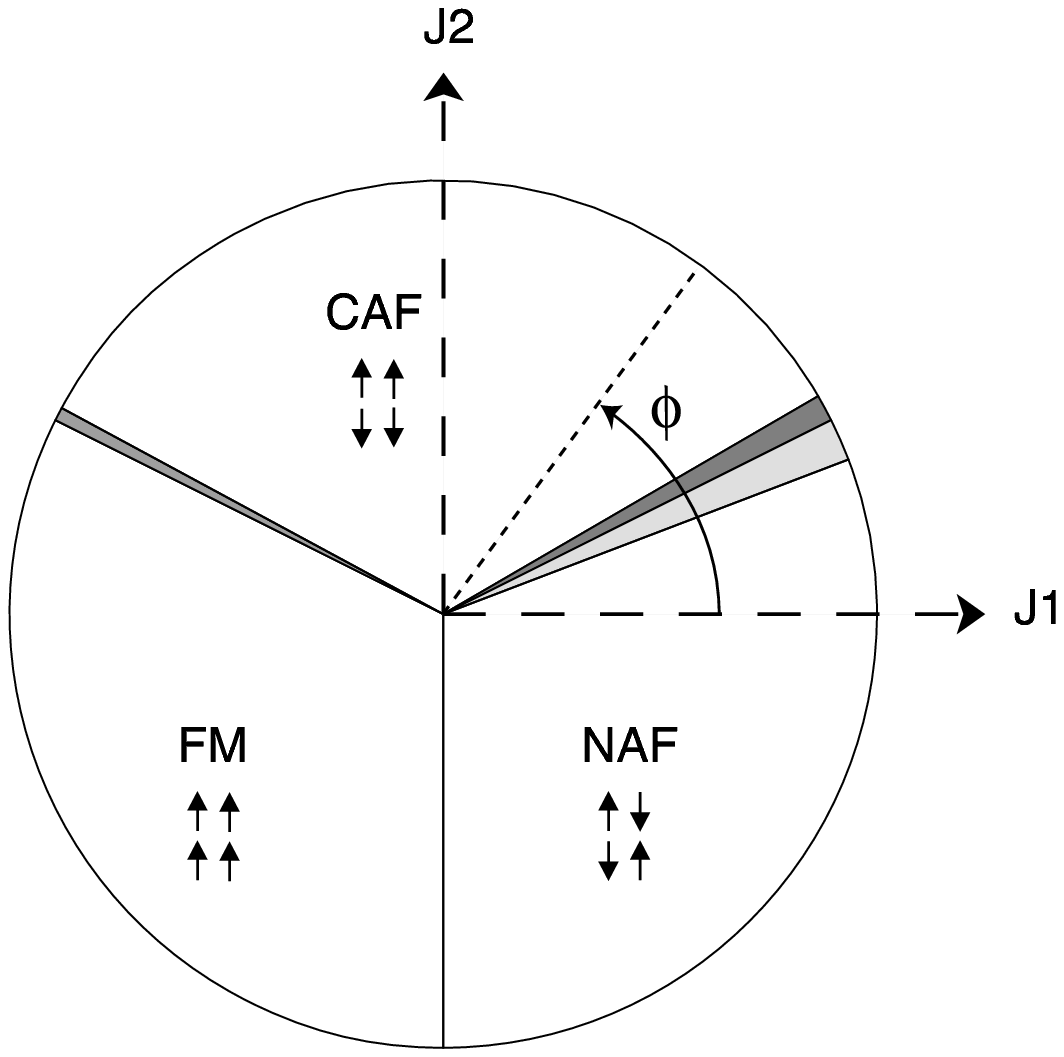}\hfill
    \includegraphics[width=.484\columnwidth]{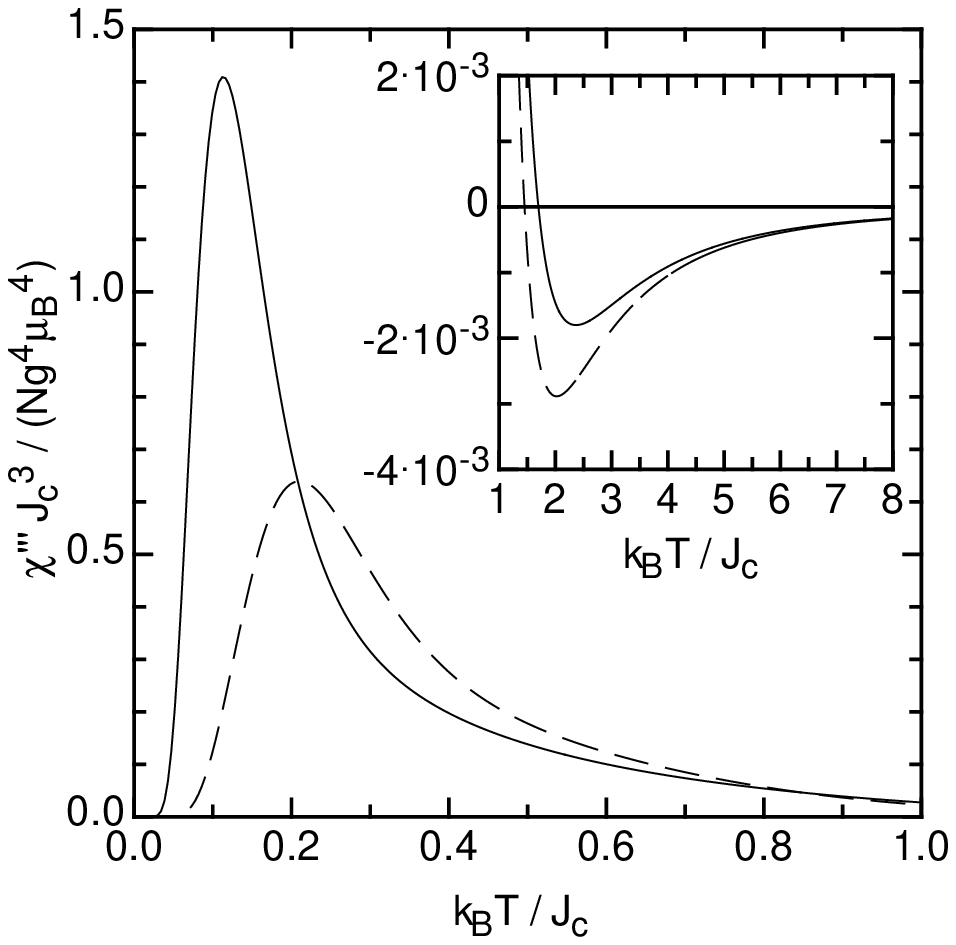}
    \caption[Phase diagram and third-order susceptibility]{Left:
    Schematic phase diagram of the frustrated Heisenberg
    antiferromagnet.  The shaded areas denote the spin-liquid region
    for $J_{1}>0$, $J_{2}\approx J_{1}/2$ and a possible new
    spin-liquid region for $J_{1}<0$, $J_{2}\approx-J_{1}/2$.\\
    Right: Temperature dependence of the third-order magnetic
    susceptibility of the frustrated Heisenberg antiferromagnet on a
    20-site cluster obtained from exact diagonalization.  The data are
    taken at two different frustration angles $\phi=-0.11\,\pi$ (solid
    line) and $\phi=0.64\,\pi$ (dashed line).  The inset shows the tiny
    minimum occurring at $T\approx2J_{\rm c}/k_{\rm B}$.}
    \label{fig:fig1}
\end{figure}

The spin-$1/2$ Heisenberg model on a two-dimensional square lattice
with next-nearest neighbour interaction belongs to the most
intensely studied models for frustrated spin systems.  Its
Hamiltonian is of the form
\begin{equation}
    H = J_1 \sum_{\langle ij \rangle_1} \vec{S}_i\cdot\vec{S}_j
    + J_2 \sum_{\langle ik \rangle_2} \vec{S}_i\cdot\vec{S}_k,
\label{eqn:H}
\end{equation}
where the sum on $\langle ij \rangle_1$ runs over nearest neighbour
and the sum $\langle ik \rangle_2$ over diagonal next-nearest
neighbour bonds.  We allow the exchange constants $J_1$ and $J_2$ to
be negative (FM) as well as positive (AF).  Only the relative size of
the exchange couplings determines the physics of the model.  Therefore
it is convenient to introduce an overall energy scale $J_{\rm
c}=\sqrt{J_{1}^2+J_{2}^2}$ and a frustration angle
$\phi=\tan^{-1}(J_2/J_1)$ to characterize the model.  The schematic
phase diagram of the model, which is shown in the left part of
Fig.~\ref{fig:fig1}, can roughly be subdivided into three ordered
phases, plus two spin-gapped phases or families of phases.  The former
are characterised by anomalies in the susceptibility $\chi({\bf q})$
at the ordering vectors ${\bf q}=(0,0)$ (FM), ${\bf q}=(\pi,0)$, or
$(0,\pi)$ (collinear antiferromagnet, CAF), and ${\bf q}=(\pi,\pi)$
(N\'{e}el-type antiferromagnet, NAF).

Experimental realisations of the frustrated square-lattice
antiferromagnet include the quasi-two-dimen\-sional compounds
Li$_2$VO(Si,Ge)O$_4$ and Pb$_2$VO(PO$_4$)$_2$~\cite{kaul:04}.  The
latter is believed to have one ferromagnetic exchange constant.  In
the following, we shall refer to this compound when discussing the
experimental situation.

In principle, the average interaction constant $J_{\rm c}$ can be
determined from the asymptotic behaviour of the heat capacity and the
magnetic susceptibility at high temperatures.  However, the
determination of the correct frustration angle, and therefore the
correct values of the exchange constants of the compounds mentioned
above has proved difficult.  Diffuse neutron scattering would provide
a means to determine $J_{1}$ and $J_{2}$ separately, therefore we have
calculated the static spin structure factor $S({\bf q},T)$
previously~\cite{shannon:04}.  Combining the parameter dependence of
the third-order susceptibility presented here with results obtained
earlier for the heat capacity and the linear susceptibility, our
method provides a way to unambiguously determine the experimental
values of $J_{1}$ and $J_{2}$ from the thermodynamic properties of the
compounds.

The third-order susceptibility $\chi'''(T)$ is defined via the 
small-field expansion of the magnetisation $M(T)$:
\begin{eqnarray}
    M 
    &=& \chi\cdot B + \frac{1}{3!}\,\chi'''\cdot B^{3} + \ldots,\\
    \chi''' &=& \frac{N_{\rm A}\mu_{0}}{N}
    \frac{(g\mu_{\rm B})^{4}}{k_{\rm B}^{3}T^{3}}
    \left(\left\langle\left(S_{\rm tot}^{z}\right)^{4}\right\rangle -
    3 \left\langle\left(S_{\rm tot}^{z}\right)^{2}\right\rangle^{2}\right),
    \label{eqn:chi3}
\end{eqnarray}
where the symbol $\langle\ldots\rangle$ denotes the trace over the
statistical operator for zero magnetic field $B$.  $S_{\rm tot}^{z}$
is the $z$ component of the total spin of the system, $N$ the system
size.  As usual, $N_{\rm A}$ is the Avogadro number, $\mu_{0}$ the
magnetic permeability, $g$ the gyromagnetic ratio, $\mu_{\rm B}$ the
Bohr magneton, and $k_{\rm B}$ the Boltzmann constant.

An expansion of the Brillouin function yields the high-temperature 
asymptotic behaviour of $\chi'''(T)$, which is of the form
\begin{eqnarray}
    \lefteqn{\chi'''(T\to\infty)=}
    \nonumber\\
    & &-\frac{N_{\rm A}\mu_{0}}{N}
    \frac{(g\mu_{\rm B})^{4}}{15}
    \cdot\frac{S(S+1)(S^{2}+(S+1)^{2})}{(k_{\rm B}T)^{3}}
    +\dots.
    \label{eq:chibrillouin}
\end{eqnarray}
Note that $\chi'''(T)$ is negative for temperatures $T\gg J_{\rm c}$.

\begin{table}[b]
    \centering
    \caption{Characteristic values for the third-order susceptibility
    obtained for the two different frustration angles $\phi=-0.11\,\pi$
    and $\phi=0.64\,\pi$ which are equally possible for the compound
    Pb$_2$VO(PO$_4$)$_2$.  The second column lists the ratio of the
    Curie-Weiss temperature to the maximum temperature of the linear
    susceptibility as obtained from
    Ref.~\protect\cite{shannon:04}.}
    
    \vspace*{\baselineskip}
    \begin{tabular}{c|cccccc}
	$\frac{\phi}{\pi}$ &
	$\frac{\Theta_{\rm CW}}{T_{\rm max}}$ &
	$\frac{k_{\rm B}T'''_{\rm max}}{J_{\rm c}}$ &
	$\frac{k_{\rm B}T'''_{0}}{J_{\rm c}}$ &
	$\frac{k_{\rm B}T'''_{\rm min}}{J_{\rm c}}$ &
	$\frac{\chi'''_{\rm max}J_{\rm c}^3}{Ng^4\mu_{\rm B}^4}$ &
	$\frac{\chi'''_{\rm min}J_{\rm c}^3}{Ng^4\mu_{\rm B}^4}$\\
	\hline
	-0.11 & 0.49 &
	0.11 & 1.69 & 2.37 & 1.41 & -0.0018\\
	\phantom{-}0.64 & 0.49 &
	0.21 & 1.45 & 2.02 & 0.64 & -0.0029\\
    \end{tabular}
    \label{tbl:tbl1}
\end{table}

We have applied the finite-temperature Lanczos method~\cite{jaklic:00}
for clusters with 16 and 20 sites to evaluate the thermodynamic
traces in Eq.~\ref{eqn:chi3}.  To give an example, the right part
of Fig.~\ref{fig:fig1} shows the temperature dependence of $\chi'''$
of a 20-site cluster for two different values $\phi_{\pm}$ of the
frustration angle.  These values are chosen such that they correspond
to those determined for
Pb$_2$VO(PO$_4$)$_2$~\cite{shannon:04,kaul:04}.  The frustration angle
$\phi=-0.11\,\pi$ (solid line) corresponds to the N\'{e}el
antiferromagnet, while $\phi=0.64\,\pi$ (dashed line) describes the
collinear phase.

\begin{figure}[t]
    \centering
    \includegraphics[width=.486\columnwidth]{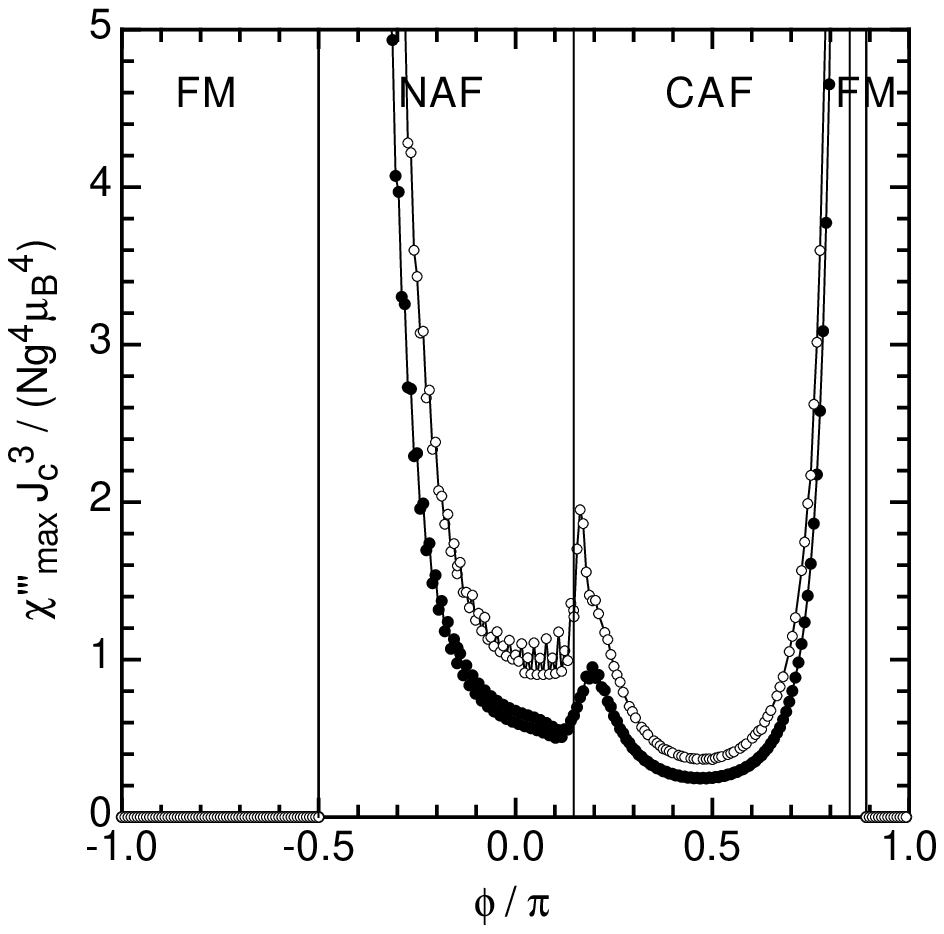}\hfill
    \includegraphics[width=.494\columnwidth]{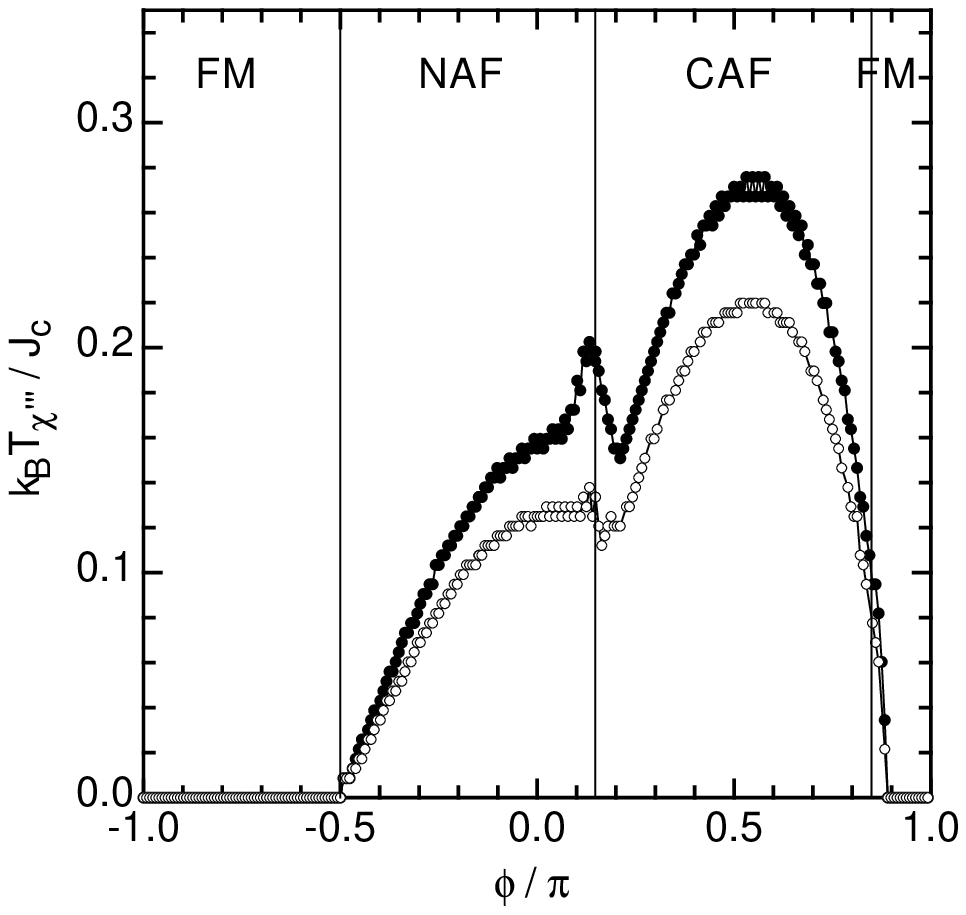}
    \caption[Maximum of the third-order susceptibility]{Maximum value
    (left) and maximum temperature (right) of the third-order magnetic
    susceptibility of the frustrated Heisenberg antiferromagnet.  The
    filled dots represent the exact-diagonalization results for a
    16-site cluster, the open circles correspond to a 20-site
    cluster.}
    \label{fig:fig2}
\end{figure}

In all non-ferromagnetic phases of the model, the temperature
dependence of $\chi'''(T)$ has a pronounced maximum at a temperature
$T_{\rm max}'''$, vanishes at a temperature $T_{0}'''$, passes through
a tiny minimum at a temperature $T_{\rm min}'''$, and eventually
approaches the high-temperature $T^{-3}$ dependence.  We have followed
these characteristic temperatures as a function of the frustration
angle.  The initial maximum temperature, together with the value of
$\chi'''$ at that point, are shown for the 16-site (solid dots) and
20-site cluster (open circles) in Fig.~\ref{fig:fig2}.  The maxima
occur at temperatures $T_{\rm max}\ll J_{\rm c}$, therefore
finite-size effects are large.

The characteristic temperatures, together with the values of $\chi'''$
at maximum and minimum, are compiled in Table~\ref{tbl:tbl1} for
$\phi=-0.11\,\pi$ and $0.64\,\pi$.  Also shown is the ratio
$\Theta_{\rm CW}/T_{\rm max}$ of the Curie-Weiss temperature to the
position of the characteristic maximum of the linear susceptibility
$\chi(T)$ taken from Ref.~\cite{shannon:04}, which is equal for
both values $\phi=\phi_{\pm}$.  Together with the results from 
Ref.~\cite{shannon:04}, our findings should be useful in
determining the precise value of $\phi$ for a given $J_{1}$-$J_{2}$
compound, as exemplified here for Pb$_2$VO(PO$_4$)$_2$.

\bibliographystyle{apsrev}
\bibliography{contrib}

\end{document}